\documentclass[letterpaper, 10 pt, conference]{ieeeconf}  % Comment this line out
\IEEEoverridecommandlockouts                              % This command is only
\usepackage[utf8]{inputenc}
\usepackage[T1]{fontenc}
\let\labelindent\relax

\usepackage{amsmath}
\usepackage{amssymb}
\usepackage{subcaption}
\usepackage{algorithmic}
\usepackage{algorithm}
\usepackage{mathtools}
\usepackage{xcolor}
\usepackage{makecell}
\usepackage{multirow}
\usepackage{graphicx}
\usepackage{marginnote}
\usepackage{float}
\usepackage{gensymb}
\usepackage{amsthm}
\newtheorem{remark}{Remark}

\newcommand{\nc}{\mathrm}

\newcommand{\cm}{\mathcal}

\usepackage[bottom]{footmisc}
\usepackage{enumitem}
\usepackage{cite}

\usepackage[scr=boondox]{mathalfa}

\title{\LARGE \bf
ALADIN-based Distributed Model Predictive Control with dynamic partitioning: An application to Solar Parabolic Trough Plants }

\author{P. Chanfreut$^{1}$, J. M. Maestre$^{2}$, D. Krishnamoorthy$^{1}$, E. F. Camacho$^{2}$% <-this % stops a space
\thanks{*This work is supported by the
European Research Council Advanced Grant OCONTSOLAR under Grant
SI-1838/24/2018, and by the Spanish MCIN/AEI/10.13039/501100011033
Project C3PO-R2D2 under Grant PID2020-119476RB-I00.}% <-this % stops a space
\thanks{$^{1}$P. Chanfreut and  D. Krishnamoorthy are with the Department of Mechanical Engineering, Eindhoven University of Technology, The Netherlands. E-mails:
        {\tt\small p.chanfreut.palacio@tue.nl, d.krishnamoorthy@tue.nl}}%
\thanks{$^{2}$ J. M. Maestre and E. F. Camacho are with the Department of Systems and Automation Engineering, University of Seville, Spain. E-mails:
        {\tt\small pepemaestre@us.es, efcamacho@us.es}}%
}

\begin{document}

\maketitle
\thispagestyle{plain}
\pagestyle{plain}

%%%%%%%%%%%%%%%%%%%%%%%%%%%%%%%%%%%%%%%%%%%%%%%%%%%%%%%%%%%%%%%%%%%%%%%%%%%%%%%%
\begin{abstract}
 This article presents a distributed model predictive controller with time-varying partitioning based on the augmented Lagrangian alternating direction inexact
Newton method (ALADIN). 
In particular, we address the problem of controlling the temperature of a heat transfer fluid (HTF) in a set of loops of solar parabolic collectors by adjusting its flow rate. The control problem involves a nonlinear prediction model,  decoupled inequality constraints, and coupled affine constraints on the system inputs. The application of~ALADIN to address such a problem is combined with a dynamic clustering-based partitioning approach that aims at reducing, with minimum performance losses, the number of variables to be coordinated. Numerical results on a~10-loop plant are~presented. 
\end{abstract}
%%%%%%%%%%%%%%%%%%%%%%%%%%%%%%%%%%%%%%%%%%%%%%%%%%%%%%%%%%%%%%%%%%%%%%%%%%%%%%%%
\section{INTRODUCTION}
%In contrast to the often used alternating direction of multipliers method 

% In Denmark, the annual share of wind and solar surpassed
% 50%, while in Ireland, Spain and Uruguay it was above 30%.
% So far, no examples exist of fully renewable-based energy
% systems that span the electricity, heating and cooling, and
% transport sectors; however, the technological, infrastructural
% and operational foundations of such systems are now being laid.
% The rise of increasingly cost-effective energy storage combined
% with greater demand-side flexibility and the expansion of
% transmission infrastructure is making it possible for regions
% with widely differing resource endowments to transition to fully
% renewable-based power systems. 

% This year’s Renewable Energy Report showcases the fact that Spain is making firm progress in the ecological transformation process, consolidating its leadership in renewable energy in 2021. Renewable energy sources in Spain accounted for nearly a 47% share of the generation mix nationwide, an all-time high made possible by the increase in electricity production using wind and solar photovoltaic technologies. Both technologies set new all-time records: wind was the leading source of energy in the mix, with a share of more than 23%, an increase of 10% year-on-year while solar photovoltaic grew by 37% and reached a share of 8% of the overall generation mix.

%% SOlar energy
%% MPC
%% Classic operation
%% DMPC
%% Coalitional
%% Main contributions
%% Outline

Over the last decades, solar energy technologies have become increasingly efficient and cost-effective, and they are now essential for the transition towards a sustainable power system. In~2021, solar power was ranked the top power
generation source installed worldwide, and has recently surpassed the threshold of~1 terawatt of installed capacity~\cite{report:SolarPowerEurope}. %Furthermore, it is projected to significantly increase during the following years as reported in \cite{report:iea_forecast}.  Solar energy technologies embrace various methods for harnessing the solar irrandiance and converting it into usable forms of energy, including photovoltaics and concentrating solar power (CSP) systems. 
 Undoubtedly, solar photovoltaics are being the kingpin for the growth of solar technologies, representing more than 50\% of the renewable  generating capacity added in 2021~\cite{report:SolarPowerEurope}. Nonetheless, there are about~6 gigawatts of concentrating solar power (CSP), and more than~1 gigawatt under construction~\cite{report:REN21}. Moreover, the incorporation of thermal energy storage technologies makes CSP systems capable of dispatching power on demand, even during the night, which is of particular interest to support other forms of renewable generation~\cite{liu2016review, alva2018overview}.

This paper focuses on solar parabolic trough plants, which represent the most extended~CSP technology~\cite{awan2020commercial}. Parabolic trough plants obtain thermal energy by concentrating the solar rays on a tube through which circulates a heat transfer fluid (HTF)~\cite{fuqiang2017progress, eddine2013parabolic}.   In this regard, the solar field  consists of a set of parallel \textit{loops}, which are rows of parabolic collectors with a tube running along their focal line. % The thermal energy gained by the HTF is then converted into mechanical energy to produce electricity. 
One of the main control problems that arise in this context is to control the HTF temperature around a given reference by manipulating its flow rate. While different control methods have been explored to address the latter, model predictive control (MPC) has received special attention both at research and commercial levels. See \cite{camacholibrosolar} (Chapter 5) for a review, and~\cite{navas2018optimal} and~\cite{gholaminejad2022stable} for recent contributions. %This paper is particularly framed within the field of distributed MPC~\cite{stewart2010cooperative},   where the decision-making power is vested in a set of interacting MPC controllers, also called agents.

Traditionally,  all loops of collectors receive the same HTF flow. However, several works have pointed out that higher efficiencies can be attained by optimally allocating the flow that circulates specifically through each loop, since they may exhibit disparate dynamics~\cite{sanchez2019thermal,  abutayeh2019effect, gallego2022nonlinear}. This is because the loops may receive different irradiance levels due to cloud shading, have different optical efficiencies due to changes in their mirrors reflectivity, etc.
%is notably influenced by the irradiance received by each loop, which may differ due to cloud shading. Further reasons are changes in their mirrors reflectivity and in the heat loss coefficients, which can vary due to dust accumulation and vacuum losses, among others. 
The associated MPC problem results in  a constrained optimization problem, where the goal is to optimally distribute the total~HTF available in the plant. The sheer size of these plants, which may comprise more than~800 loops as in~SOLANA~\cite{SOLANA}, hinders the applicability of a centralized MPC approach. % due to its strong computational and communication demands. 
 Pursuing increased scalability, a number of articles have explored distributed MPC~(DMPC) strategies where multiple agents control subsets of loops, e.g.,~\cite{frejo2020centralized,  sanchez2023coalitional}. In addition, DMPC is also favourable in terms of monitoring and maintenance. For example, if some of the loops are not operating, it will only affect some of the agents, while the rest could continue operating normally. 

Within the DMPC framework, dual decomposition and the alternating direction method of multipliers (ADMM) have been extensively used for coordinating control decisions~\cite{giselsson2013accelerated}. Both of them involve iterative procedures based on (sub)gradient methods, which often require many iterations before converging to a solution. Moreover, their theoretical properties do not generally apply in the nonconvex setting~\cite{houska2016augmented}. Considering these issues, this paper explores the augmented Lagragian alternating direction inexact Newton method (ALADIN)~\cite{houska2016augmented, engelmann2022aladinalpha}, which has been recently studied for optimizing power transfers in electrical networks~\cite{engelmann2018toward, jiang2020distributed}. Particularly, ALADIN combines ideas of augmented Lagrangian methods and sequential quadratic programming, and is designed to solve potentially nonconvex optimization problems in a distributed manner. In contrast to ADMM, ALADIN uses both gradient and Hessian information at every iteration, and has been shown to converge faster~\cite{engelmann2022aladinalpha}. This is beneficial for the real-time control problem underlying our solar plants application.

 The main contribution of this article is a DMPC based on ALADIN with time-varying system partitioning. The proposed controller optimizes the HTF  flow rates in every loop to track reference outlet temperatures, and integrates clustering methods to further increase scalability with minimum performance losses.  In this regard, the solar field is dynamically partitioned into clusters of similar loops to reduce the number of optimization variables, and thus simplify the distributed computations.  %Our results on a 10-loop plant  show that the proposed approach can closely approximate the performance of DMPC approaches based on static finer partitions. 

The rest of the article is organized as follows. Section~\ref{sec:problem_formulation} presents the system dynamics, the control objectives, and the associated centralized problem. Section~\ref{sec:Clust_aladin} describes the clustering formation, formulates the DMPC problem, and presents the proposed ALADIN-based control algorithm. Finally, Section~\ref{sec:simulations} presents our simulation results. %, and Section~\ref{sec:conclusions} concludes the paper. 

\vspace{4pt}
\textit{Notation:} Given two time steps $k$ and $n\geq k$, and a variable~$x$, $x(n|k)$ indicates the predicted value of $x$ for time~$n$ realized at~$k$. Given a set $\cm{S}$, say $\cm{S} = \{1,2,...,|\cm{S}|\}$, $[x_i]_{i\in \cm{S}}=[x_i]_{i=1}^{|\cm{S}|}$ is the vector $[x_1, x_2, ..., x_{|\cm{S}|}]^\top$. Also, $|\cdot|$ denotes the cardinality when referring to a set, and the absolute value when used with scalars. Capital caligraphic letters are used for sets, whereas bold letters represent sequences. Finally,~$\mathbf{1}_{m}$ and~$\mathbf{0}_{m}$ are the all-ones and all-zeros vectors of dimension $m\times 1$.

\section{Problem formulation}~\label{sec:problem_formulation}
\vspace{2pt}
Consider a solar parabolic trough plant comprising a set of parallel loops $\cm{N}=\{1, 2, ..., N_\nc{loops}\}$ equipped with inlet valves (see Fig.~\ref{fig:dibujo_solar_plant}). 

\subsection{System dynamics}\label{subsec:system_dynamics}

The dynamics of the HTF temperature at the outlet of any loop $i \in \cm{N}$, i.e., $T^\nc{out}_i$ [\degree C], can be modeled  considering the variation of its internal energy as follows:\footnote{For the sake of clarity, the continuous time index is omitted in Subsections~\ref{subsec:system_dynamics} and~\ref{subsec:control_objectives}.}
\begin{equation}\label{eq:loops_model}
    C_i \dfrac{d T^\nc{out}_i}{dt}  = \eta_i \mathscr{I}_i - q_i P_i (T^\nc{out}_i - T^\nc{in}) - \mathscr{h}_i,
\end{equation}
%Kopt=Reflectividad*Abstubo*Factordeforma*Transmitancia*malapuntamiento
%intercept factor (γ) which is defined as the fraction of rays incident upon the aperture that reach the receiver for a given incidence angle. 
%The optical efficiency η (dimensionless) of a receiver tube is defined as the fraction of the incident solar radiation energy on the collectors aperture which is transferred to a heat transfer fluid as thermal energy inside the absorber tube
\noindent where $T^\nc{in}$ [\degree C] is the inlet temperature, and $q_i$ [m$^3$/s] represents the~HTF flow rate in loop $i$. Also, $C_i$ [J/ºC] is the thermal capacity of the loop, $P_i$ [J/(m$^3$\degree C)] is related to its geometrical and thermal properties, $\mathscr{h}_i$ [W] is a function weighting the heat losses of loop $i$, and $\eta_i \mathscr{I}_i$ [W] considers the power received from the sun. In particular, $\eta_i$ weights the optical and geometric efficiency of the collectors in $i$, and $\mathscr{I}_i = S I_i$, with~{$S$~[m$^2$]} being the loops' reflective surface  and~$I_i$~[W/m$^2$] the direct normal irradiance. %This parameter takes into account the mirrors reflectivity, the interception factor, the absorptance of the tube, and the position of the collectors with respect to the radiation beam vector, among others~\citep{YILMAZ2018135,wei2020simplified}. 
Finally, note that some of the parameters in model~\eqref{eq:loops_model} vary as a function of the temperature. In particular, we will consider the following throughout this paper\footnote{The definitions in~\eqref{eq:parameters_loops} consider the HTF (Therminol~55) and heat losses of the ACUREX
plant, which is located in the south of Spain~\cite{camacholibrosolar}.}:
\begin{equation}\label{eq:parameters_loops}
\begin{split}
&\rho_i  =  903 - 0.672 T^\nc{m}_i, \qquad \ \ P_i = \rho_i c_i, \\
&  c_i = 1820 + 3.478 T^\nc{m}_i,  \qquad  C_i = \rho_i c_i A L,   \\
  &  \mathscr{h}_i = S \left(0.00249(T^\nc{m}_i  - T^\nc{a})^2 - 0.06133 (T^\nc{m}_i  - T^\nc{a}) \right), 
    \end{split}
\end{equation}
\noindent where $T^\nc{m}_i = (T^\nc{out}_i + T^\nc{in})/{2}$  is the mean between the inlet and outlet temperature of loop $i$, 
$T^\nc{a}$ [\degree C] is the ambient temperature,~$A$ [m$^2$] is the cross sectional area of the tube, and~$L$~[m] is the loops length.

\subsubsection{Cluster-based model}\label{sec:cluster_based_model}
Similar to \eqref{eq:loops_model}, a cluster of loops $\cm{C} \subseteq \cm{N}$ can be jointly described by the following lumped parameter model:
\begin{equation}\label{eq:clusters_model}
    C_{\cm{C}} \dfrac{d T^\nc{out}_{\cm{C}}}{dt}  =  \eta_{\cm{C}}\mathscr{I}_{\cm{C}} - q_{\cm{C}} P_{\cm{C}} (T^\nc{out}_{\cm{C}} - T^\nc{in}) - \mathscr{h}_{\cm{C}},
\end{equation}
where $T^\nc{out}_{\cm{C}}$ denotes the outlet temperature of cluster $\cm{C}$, and~$q_{\cm{C}}$ is the total HTF pumped to the loops in~$\cm{C}$. Also, parameters  $C_{\cm{C}}$,  $P_{\cm{C}}$ and $\mathscr{h}_{\cm{C}}$ are defined analogously to~\eqref{eq:parameters_loops}, and $ \eta_{\cm{C}}\mathscr{I}_{\cm{C}} = \sum_{i \in \cm{C}} \eta_i \mathscr{I}_i$. Note that if $\cm{C}=\cm{N}$, then~\eqref{eq:clusters_model} provides a lumped model of the entire solar~field; whereas if $\cm{C}=\{i\}$, model~\eqref{eq:clusters_model} is equivalent to~\eqref{eq:loops_model}.

% and the thermal losses are computed as: 
% \begin{equation}
%    \mathscr{h}_{\cm{C}_j} =  |\cm{C}_j| S \left( h_{2,\cm{C}_j} (T^\nc{m}_{\cm{C}_j}  - T^\nc{a})^2 + h_{1,\cm{C}_j} (T^\nc{m}_{\cm{C}_j}  - T^\nc{a}) \right), 
% \end{equation}
% \noindent  where $|\cm{C}_j|$ is the cardinality of $\cm{C}_j$, i.e., the number of loops in cluster $j$. 

 \subsection{Control objectives}\label{subsec:control_objectives}

%Hereafter, consider a distributed setting where each loop represents a subsystem that is controlled by an MPC agent. 
The proposed controller should dynamically update flow rates $q_i$ for all $i \in \cm{N}$ so as to track time-varying references on the loops outlet temperature while satisfying the following constraints:
 \begin{subequations}\label{eq:constraints_q}
      \begin{align}
     &\sum_{i\in \cm{N}} q_i \leq Q_\nc{T}, \label{eq:coupled_constraint} \\  \ \ &q^\nc{min} \leq q_i \leq q^\nc{max}, \ \forall i \in \cm{N}, \label{eq:local_constraint} \\
     & T^\nc{min} \leq T_i^\nc{out} \leq T^\nc{max}, \ \forall i \in \cm{N},
     \end{align}
      \end{subequations}
 \noindent where $Q_\nc{T}$ is the maximum available HTF flow in the plant, $q_\nc{min}$ and $q_\nc{max}$ denote respectively the minimum and maximum flows  allowed in the loops, and $T^\nc{min}$ and $ T^\nc{max}$ are similarly the minimum and maximum desired temperatures. Note that, as long as~\eqref{eq:coupled_constraint} is satisfied, the total available HTF can be unevenly distributed among the set of loops, e.g., higher flow rates can be pumped to loops receiving greater irradiance. Finally, the proposed controller should be scalable and approximate the optimal performance with reduced computational and communication burden.

%to minimize the mismatch between loops the prediction models and their dynamics

 % Moreover, we will two different approaches with regard to the reference temperatures. First, we will consider an (admissible) arbitrary piecewise reference, while, secondly, this reference will be updated to maximize the net energy. 
 
 \subsection{Centralized MPC problem}

 In what follows, consider a discrete-time setting, let~$\Delta t^\nc{s}$ be the integration step size, and let~$k$ be the discrete time index, i.e., step $k$ refers to instant $k\Delta t^\nc{s}$.  Likewise, let~$\Delta t^\nc{c}= \delta^\nc{c} \Delta t^\nc{s}$ be the sampling time considered in the control models, where~$\delta^\nc{c} \in~\mathbb{N}^+$. Then, the centralized MPC problem underlying this article can be formulated as follows:
\begin{subequations}\label{eq:cen_problem}
\begin{equation*}
    \min_{[\mathbf{q}_i(k)]_{i\in \cm{N}}} \ \ \  \sum_{i \in \cm{N}}  \ \sum_{n \in \cm{H}}  \ \Big( w_\nc{e} e_i^2(n + \delta^\nc{c}|k)  +  w_\nc{q} q_i^2(n|k) \Big) 
\end{equation*}
\vspace{-10pt}
\begin{align}
    &\text{s.t.}  \nonumber \\  
     \begin{split} &T^\nc{out}_i(n+\delta^\nc{c}|k)   = T^\nc{out}_i(n|k) \\ & \quad  + \dfrac{\Delta t^\nc{c}}{C_i(n|k)} \Big( \eta_i(k)\mathscr{I}_i(k) -   \mathscr{h}_i(n|k) \Big)\\ &\quad  -  \dfrac{\Delta t^\nc{c}}{C_i(n|k)} q_i(n|k) P_i(n|k) \left(T^\nc{out}_i(n|k) - T^\nc{in}(k)\right), \end{split} \label{eq:model_opt_problem_bottom} \\[2pt]
    & T^\nc{out}_i(k|k) = T^\nc{out}_i(k),\\ 
    & T^\nc{min} \leq T^\nc{out}_i (n+ \delta^\nc{c}|k) \leq T^\nc{max}, \label{eq:constr_T}\\
    & q^\nc{min} \leq q_i(n|k) \leq q^\nc{max}, \\
    & \sum_{l \in \cm{N}} q_l(n|k) \leq  Q_\nc{T}, \label{eq:alpha_loops}\\
    & \forall i \in \cm{N}, \ \forall n \in \cm{H},
\end{align}
\end{subequations}
\noindent where $e_i(n+\delta^\nc{c}|k) = T^\nc{out}_i(n\!+\!\delta^\nc{c}|k) - T^\nc{ref}(n\!+\!\delta^\nc{c})$ denotes the outlet temperature error of loop $i$, with $T^\nc{ref}(\cdot)$ being the reference temperature. Also, $\cm{H}=\{k, k+\delta^\nc{c},  k+2\delta^\nc{c}, ..., k+ \delta^\nc{c} N_\nc{p}\}$ is the set of time instants considered in the prediction horizon, with $N_\nc{p}$ being a tuning parameter, $\mathbf{q}_i(k) = [q_i(k|k), q_i(k+ \delta^\nc{c}|k), ..., k+\delta^\nc{c}N_\nc{p}]^\top$ is the flow rate sequence of loop $i$, and~$w_\nc{e}$ and~$w_\nc{q}$ are positive definite weighting scalars. 
 Likewise,~\eqref{eq:model_opt_problem_bottom} is a discrete-time version of model~\eqref{eq:loops_model}, where~$P_i(n|k)$,  $C_i(n|k)$ and $\mathscr{h}_i(n|k)$ are computed considering~\eqref{eq:parameters_loops} and the predicted mean temperature   $T_i^\nc{m}(n|k) = (T^\nc{out}_i(n|k) + T^\nc{in}(k))/2$. 
Finally, for the sake of simplicity, the inlet temperature,  effective irradiance, and ambient temperature, are assumed to mantain their value at~$k$ during the entire prediction horizon.
 
\begin{remark}
Prediction model~\eqref{eq:model_opt_problem_bottom} introduces nonconvex terms both in the cost function and in constraint~\eqref{eq:constr_T}.  %Note that $T^\nc{out}_i(\cdot|k)$ multiplies~$q_i(\cdot|k)$, and thus introduces bilinear terms.
 Therefore, it is not possible to claim that~\eqref{eq:cen_problem} is generally a convex optimization problem.    %(see~\eqref{eq:parameters_loops}).
\end{remark}

%Overall, the approach aims at providing a scalable solution to tackle the flows optimization problem given by the optimal distribution of $Q_\nc{T}$ among all the loops.
%. As can be seen, it consists of a set of distributed MPC agents that dynamically interact with a supervisory layer. 

%Note that \eqref{eq:coupled_constraint} couples all flow rate variables. 

\section{Clustering-based DMPC using ALADIN}~\label{sec:Clust_aladin}
\begin{figure}[t]
    \centering
    \includegraphics[scale=0.38, trim={3.7cm 4.2cm 2.9cm 3cm},clip]{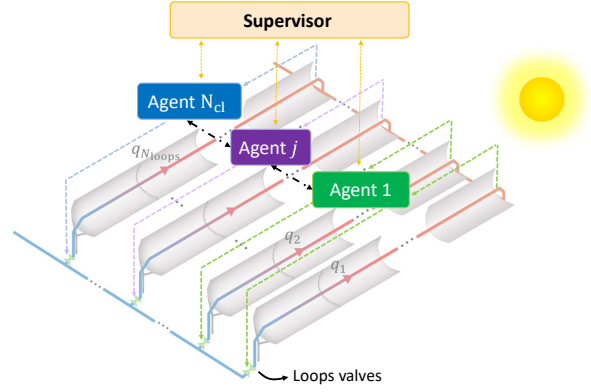}
    \caption{Architecture of the proposed control approach. The agents control the flow rates in different clusters of loops, e.g., agent 1 controls loops 1 and 2.}
    \label{fig:dibujo_solar_plant}
\end{figure}

Centralized problem~\eqref{eq:cen_problem} may involve a large number of loops and lacks convexity guarantees as mentioned above. Considering this issue, this article proposes the distributed control architecture illustrated in  Fig.~\ref{fig:dibujo_solar_plant}, which comprises a set of MPC agents and a supervisor. The main features of this approach are the following:\vspace{3pt}
\begin{itemize}[noitemsep]
\item[(i)] The set of $N_\nc{loops}$ loops are \textit{dynamically} partitioned by the supervisor  into a set of non-overlapping clusters $\{ \cm{C}_1, \cm{C}_2, ..., \cm{C}_{N_\nc{cl}}\}$, such that
\begin{equation*}
    \bigcup_{j=1}^{N_\nc{cl}} \cm{C}_j = \cm{N}, \ \ \text{and} \  \ \cm{C}_j \cap \cm{C}_l = \emptyset \ \  \forall j,l \in [1, N_\nc{cl}], \ j \neq l, 
\end{equation*}
\noindent where $N_\nc{cl}\leq N_\nc{loops}$ denotes the number of clusters.
\item[(ii)] Each resulting cluster $\cm{C}_j$ is assigned to MPC agent~$j$, which controls flow rates $q_i$ for all $i \in \cm{C}_j$ during a given time period. 
\item[(iii)] The set of MPC agents coordinate their decisions to optimize their collective performance using ALADIN, which is designed to address (potentially nonconvex) distributed problems.
\end{itemize}

%   \begin{figure}[b]
%     \centering
%     \includegraphics[scale=0.38, trim={2.5cm 4cm 1.5cm 2.5cm},clip]{figs/fig_architecture.pdf}
%     \caption{Hierarchical architecture of the proposed approach. At the bottom layer, a set of agents control the HTF flow rates in each loop, whereas the top layer adjusts their level of coordination and constraints considering~\eqref{eq:constraints_q}.}
%     \label{fig:dibujo_architecture}
% \end{figure}

The next subsections provide further details regarding the partition selection and the proposed control algorithm. 
%This section focuses first on the partition selection, and subsequently introduces the underlying DMPC optimization and the proposed control algorithm. 

\subsection{Partition selection}\label{sec:cluster_formation}
 
 Inspired by~\cite{chanfreut2023clustering}, our proposed DMPC approach  exploits similarities between the loops to reduce the control problem complexity. In particular, the solar field is dynamically partitioned into clusters of loops whose dynamics are approximately characterized by the same parameters.  To this end,   mean temperature $T_i^\nc{m}(k)$ and current effective irradiance~$\eta_i(k) \mathscr{I}_i(k)$ are periodically collected for all $i\in \cm{N}$ so that we build the following data set:
\begin{equation}
\mathcal{D}(k)= \{ [\eta_i(k) \mathscr{I}_i(k),  \ \ T_i^\nc{m}(k)]\}_{i \in \cm{N}}.
\end{equation}
Note that, given~\eqref{eq:model_opt_problem_bottom},  those loops for which these two features are equal will have identical prediction models.  
Using clustering methods~\cite{xu2005survey}, the loops in $\cm{N}$ can then be partitioned into a number of clusters, say $N_\nc{cl}(k)\leq N_\nc{cl}^\nc{max}$, according to the data in $\mathcal{D}(k)$. In this respect, $N_\nc{cl}^\nc{max}$ denotes the maximum number of clusters, which is directly related with the number of MPC agents available in the system.  
  Without loss of generality, we consider the well-known centroid-based algorithm $K$-means~\cite{ahmed2020k}, together with the \textit{elbow} method to select the optimal number of clusters. Note that the elbow method runs the $K$-means algorithm for different $N_\nc{cl}(k)$ and computes in each case an average score evaluating the resulting partition. 

% .  In this regard, we consider 
%  the \textit{elbow} method to find the optimal $N_\nc{cl}(k)$ in range~$[2,N_\nc{cl}^\nc{max}]$, where $N_\nc{cl}^\nc{max} \leq N_\nc{cl}^\nc{max}$ is a tuning parameter denoting the maximum allowed number of clusters. 

\subsection{Clusters-based MPC problem}\label{sec:inner_cluster}
 
Let~$\cm{P}(k) = \{\cm{C}_1, \cm{C}_2, ..., \cm{C}_{N_\nc{cl}(k)}\}$ be the partition selected at time $k$ as described above. % and let each cluster $\cm{C}_j$ be controlled by agent $j$, i.e., agent~$j$ receives data from all loops $i\in \cm{C}_j$ and manipulates their flow rates. 
%In this regard, each agent considers a lumped parameter model of its cluster based on~\eqref{eq:clusters_model}, which provides a simplified representation of the corresponding set of loops. Considering this, the set of MPC agents coordinate the clusters flow rates to optimize the overall performance while satisfying constraints~\eqref{eq:constraints_q}. 
Then, we consider the following MPC problem to find the~HTF to be pumped to every cluster: 
%by all agents $j \in \{1,2,...,N_{\nc{cl}}(k)\}$:
\begin{subequations}\label{eq:clusters_optimization}
\begin{equation*} %\resizebox{1\linewidth}{!}{ $
   \min_{[\mathbf{q}_{\cm{C}_j}(k)]_{\forall \cm{C}_j}}  \sum_{j=1}^{N_\nc{cl}(k)}  |\cm{C}_j| 
 \sum_{n \in \cm{H}} \!\! \big( w_\nc{e} e_{\cm{C}_j}^2(n\!+\! \delta^\nc{c}|k) +  w_\nc{q} q_{\cm{C}_j}^2(n|k) \big) % }
\end{equation*}
\vspace{-9pt}
\begin{align}
    &\text{s.t.} \nonumber \\  
    \begin{split}
        & T^\nc{out}_{\cm{C}_j}(n+\delta^\nc{c}|k)  = T^\nc{out}_{\cm{C}_j}(n|k) \\ &+ \dfrac{\Delta t^\nc{c}}{C_{\cm{C}_j}(n|k)} \big(\eta_{\cm{C}_j}(k) \mathscr{I}_{\cm{C}_j}(k) - \mathscr{h}_{\cm{C}_j}(n|k) \big)
         \\ 
         & - \dfrac{\Delta t^\nc{c}}{C_{\cm{C}_j}(n|k)}  q_{\cm{C}_j}(n|k) P_{\cm{C}_j}(n|k) \left(T^\nc{out}_{\cm{C}_j}(n|k) - T^\nc{in}(n|k)\right)\!,
        \end{split} \label{eq:pred_model_cluster}\\
    & T^\nc{out}_{\cm{C}_j}(k|k) = \dfrac{\sum_{i \in \cm{C}_j} (q_i(k-1) T^\nc{out}_i(k))}{\sum_{i \in \cm{C}_j} q_i(k-1)},\\
    & T^\nc{min} \leq T^\nc{out}_{\cm{C}_j} (n+ \delta^\nc{c}|k) \leq T^\nc{max}, \label{eq:constT_clust}\\
    & q^\nc{min}_{\cm{C}_j} \leq q_{\cm{C}_j}(n|k) \leq  q^\nc{max}_{\cm{C}_j}, \\
    %& \delta^\nc{top} \Delta q^\nc{min} \leq \Delta q_{\cm{C}_j}(n|k) \leq  \delta^\nc{top} \Delta q^\nc{max}, \\
    & \sum_{l=1}^{N_\nc{cl}(k)} q_{\cm{C}_l}(n|k) \leq  Q_\nc{T}, \label{eq:constr_total_flow}\\
    & \forall j \in \{1, 2, ..., N_\nc{cl}(k)\},  \ \forall n \in \cm{H},
\end{align}
\end{subequations}
\noindent where the predicted outlet temperature error of the $j$-th cluster is  $e_{\cm{C}_j}(n+\delta^\nc{c}|k) = T^\nc{out}_{\cm{C}_j}(n\!+\!\delta^\nc{c}|k) - T^\nc{ref}(n\!+\!1)$, for $n \in \cm{H}$. Also, 
$q^\nc{min}_{\cm{C}_j}= |\cm{C}_j| q^\nc{min}$, $q^\nc{max}_{\cm{C}_j}=|\cm{C}_j| q^\nc{max}$, and
$$\mathbf{q}_{\cm{C}_j}(k) = [q_{\cm{C}_j}(n|k)]_{n\in\cm{H}} =[q_{\cm{C}_j}(k|k), ..., q_{\cm{C}_j}(k+\delta^\nc{c}N_\nc{p}|k)]^\top. 
$$
%Finally, notice that an uniform distribution of the maximum flow implies~$\alpha_{\cm{C}_j}(n) = |\cm{C}_j|/N_\nc{loops}$ for all $n$, whereas this approach introduces further flexibility.
\noindent Given the solution of~\eqref{eq:clusters_optimization}, say $\mathbf{q}_{\cm{C}_j}^*(k)$ for all $\cm{C}_j\in\cm{P}(k)$, the HTF is uniformly distributed among the loops in every cluster. That is, the implemented flows are given by   
\begin{equation}\label{eq:alloc}
    q_i(t) = \dfrac{q_{\cm{C}_j}^*(k|k)}{|\cm{C}_j|},  \ \forall i \in \cm{C}_j, \ \forall t \in [k, k+1,..., k +\delta^\nc{c}).
\end{equation}

\begin{remark}
Problem~\eqref{eq:clusters_optimization} has the same form as problem~\eqref{eq:cen_problem} but involves a reduced number of optimization variables. In particular, while the number of flow variables in \eqref{eq:cen_problem} was $N_\nc{p}N_\nc{loops}$, here we deal with  $N_\nc{p}N_\nc{cl}(k)$. %Note that~\eqref{eq:clusters_optimization} considers a lumped parameter model of every cluster, and thus a simplified representation of the solar field.
\end{remark}
\begin{remark}
    Since the clusters are chosen to aggregate loops with similar dynamics, the solution of~\eqref{eq:clusters_optimization} will approximate that of~\eqref{eq:cen_problem}. Particularly, we are replacing models of loops that are nearly identical with a single lumped description. This similarity among loops also motivates the uniform flow allocation indicated in~\eqref{eq:alloc}. 
\end{remark}

\subsubsection{Formulation to use ALADIN}
 As detailed in~\cite{houska2016augmented},  ALADIN is designed to solve optimization problems with separable (potentially nonconvex) objective functions, decoupled inequality constraints, and coupled affine equality constraints.

Note that, by definition, the objective function in~\eqref{eq:clusters_optimization} is separable and can indeed be rewritten as  
 % &\sum_{\cm{C}_j \in \cm{P}(k)}  f_{ \cm{C}_j}(\mathbf{e}_{\cm{C}_j}(k), \mathbf{q}_{\cm{C}_j}(k)) =
\begin{equation*}
\begin{split}
    %J_{\cm{N}}&([\mathbf{e}_{\cm{C}_j}(k), \mathbf{q}_{\cm{C}_j}(k)]_{\cm{C}_j \in \cm{P}(k)}) = \\[4pt]
     %&
     \sum_{j=1}^{N_\nc{cl}(k)}  \underbrace{|\cm{C}_j| \sum_{n \in \cm{H}}(w_\nc{e} e_{\cm{C}_j}^2(n\!+\!\delta^\nc{c}|k) + w_\nc{q} q_{\cm{C}_j}^2(n|k)).}_{f_{ \cm{C}_j}(\mathbf{e}_{\cm{C}_j}(k), 
 \mathbf{q}_{\cm{C}_j}(k))}
    \end{split}
\end{equation*}
 % \begin{equation*}%\resizebox{1.01\columnwidth}{!}{$
 %     f_{\cm{C}_j}(\cdot) = |\cm{C}_j| \sum_{n \in \cm{H}}  \left( w_\nc{e} e_{\cm{C}_j}^2(n\!+\!\delta^\nc{c}|k) + w_\nc{q} q_{\cm{C}_j}^2(n|k) \right)\!\!. %$}
 % \end{equation*}
 Likewise, given~\eqref{eq:pred_model_cluster}, variables $T_{\cm{C}_j}^\nc{out}(k+\kappa\delta^\nc{c}|k)$ for any $\kappa \in \{1,2,..., N_\nc{p}\}$ can be computed as
 \begin{equation*}\label{eq:T_qseq}
 \begin{split}
         & T^\nc{out}_{\cm{C}_j}(k+\kappa\delta^\nc{c}|k)  =  T^\nc{out}_{\cm{C}_j}(k) \ \  \\ &+ \!\!\!\sum_{\tilde{n}=k}^{k+(\kappa-1)\delta^\nc{c}} \!\!\!\!\!\!\dfrac{\Delta t^\nc{c}}{C_{\cm{C}_j}(\tilde{n}|k)} \Big( \eta_{\cm{C}_j}(k)\mathscr{I}_{\cm{C}_j}(k) -   \mathscr{h}_{\cm{C}_j}(\tilde{n}|k) \Big)\\ & - \!\!\! \sum_{\tilde{n}=k}^{k+(\kappa-1)\delta^\nc{c}}\!\!\!\!\!\!\dfrac{\Delta t^\nc{c}}{C_{\cm{C}_j}(\tilde{n}|k)}  q_{\cm{C}_j}(\tilde{n}|k) P_{\cm{C}_j}(\tilde{n}|k) \left(T^\nc{out}_{\cm{C}_j}(\tilde{n}|k) - T^\nc{in}(k)\right)\!\!.
 \end{split}
 \end{equation*}
 \noindent That is, they are a function of current outlet temperature~$T_{\cm{C}_j}^\nc{out}(k)$, inlet temperature $T^\nc{in}(k)$, effective irradiance $\eta_{\cm{C}_j}(k)\mathscr{I}_{\cm{C}_j}(k)$, ambient temperature~$T^\nc{a}(k)$, and the sequence of control inputs implemented up to time instant $k+(\kappa-1) \delta^\nc{c}$. For simplicity, let us define $z_{\cm{C}_j}(k) = [T_{\cm{C}_j}^\nc{out}(k), \eta_{\cm{C}_j}(k)\mathscr{I}_{\cm{C}_j}(k), T^\nc{in}(k), T^\nc{a}(k)]$ and $\mathbf{T}^\nc{ref}(k) = [T^\nc{ref}(n+\delta^\nc{c})]_{n\in \cm{H}}$. Then, %considering~\eqref{eq:T_qseq} and the expression of  $e_{\cm{C}_j}(n+\delta^\nc{c})$ for all $n \in \cm{H}$,  
 the objective function in~\eqref{eq:clusters_optimization} can also be rewritten as
% \begin{equation}%\resizebox{0.99\hsize}{!}{%
%     J_{\cm{N}}(z(k), \mathbf{q}(k), \mathbf{T}^\nc{ref}(k)) = \sum_{{\cm{C}_j}\in \cm{P}(k)}  f_{\cm{C}_j}(z_{\cm{C}_j}(k), \mathbf{T}^\nc{ref}(k), \mathbf{q}_{\cm{C}_j}(k)),%$}
% \end{equation}
\begin{equation}%\resizebox{0.99\hsize}{!}{%
    \sum_{j=1}^{N_\nc{cl}(k)}   f_{\cm{C}_j}(z_{\cm{C}_j}(k), \mathbf{T}^\nc{ref}(k), \mathbf{q}_{\cm{C}_j}(k)).%$}
\end{equation}
%where $z(k) = [z_{\cm{C}_j}(k)]_{{\cm{C}_j} \in \cm{P}(k)}$. 
Using the same reasoning, constraint~\eqref{eq:constT_clust} is of the form $h_{\cm{C}_j}(z_{\cm{C}_j}(k), \mathbf{q}_{\cm{C}_j}(k)) \leq \mathbf{0}_{N_\nc{p}}$, where $h_{\cm{C}_j}(\cdot): \mathbb R \times \mathbb R^{N_\nc{p}} \rightarrow \mathbb R^{N_\nc{p}}$ is the corresponding constraint function. % which considers \eqref{eq:T_qseq} for all $\kappa \in [1,..., N_\nc{p}]$.  
%Finally, considering that $T_i^\nc{out}(k)$ and $\mathbf{T}^\nc{ref}(k)$ enter the problem as parameters, and defining $z_i(k) = [T_i^\nc{out}(k), \mathbf{T}^\nc{ref}(k)]$, 
% Therefore, optimization problem \eqref{eq:clusters_optimization} can be equivalently formulated as:
% \begin{subequations}\label{eq:loops_problem_2}
% \begin{equation*}
%     \min_{[\mathbf{q}_{\cm{C}_j}(k)]_{\forall \cm{C}_j}}   \sum_{{\cm{C}_j} \in \cm{P}(k)}  f_{\cm{C}_j}(z_{\cm{C}_j}(k), \mathbf{T}^\nc{ref}(k),  \mathbf{q}_{\cm{C}_j}(k)) \qquad \qquad \qquad \qquad
% \end{equation*}
% \vspace{-10pt}
% \begin{align}
%     \text{s.t.} \ \ 
%     & h_{\cm{C}_j}(z_{\cm{C}_j}(k), \mathbf{q}_{\cm{C}_j}(k)) \leq 0, \ \  \forall {\cm{C}_j} \in \cm{P}(k), \label{eq:constr_T_2}\\
%     & \mathbf{q}_{\cm{C}_j}(k) \leq |\cm{C}_j| q^\nc{max} \mathbf{1}_{N_\nc{p}}, \ \  \forall {\cm{C}_j} \in \cm{P}(k), \\
%      -&\mathbf{q}_{\cm{C}_j}(k) \leq -|\cm{C}_j| q^\nc{min} \mathbf{1}_{N_\nc{p}}, \hspace{2pt}  \forall {\cm{C}_j} \in \cm{P}(k), \\
%     & \sum_{{\cm{C}_j} \in \cm{P}(k)} \mathbf{q}_{\cm{C}_j}(k) \leq  Q_\nc{T} \mathbf{1}_{N_\nc{p}}. \label{eq:alpha_loops_2}
% \end{align}
% \end{subequations}
\noindent Finally, let us introduce a \textit{sink} artificial loop, say loop 0, and let us define~$\cm{C}_0=\{0\}$ to keep the notation simple. Then, 
 problem~\eqref{eq:clusters_optimization} can be reformulated~as follows: 
\begin{subequations}\label{eq:loops_problem_3}
\begin{equation*}
    \min_{[\mathbf{q}_{\cm{C}_j}(k)]_{j=0}^{N_\nc{cl}(k)}}   \sum_{j=1}^{N_\nc{cl}(k)} \!\! f_{\cm{C}_j}(z_{\cm{C}_j}(k), \mathbf{T}^\nc{ref}(k),  \mathbf{q}_{\cm{C}_j}(k)) + f_{\cm{C}_0}(\mathbf{q}_{\cm{C}_0}(k))) \qquad \qquad
\end{equation*}
\vspace{-10pt}
\begin{align}
    \text{s.t.} \ \ \ 
    & h_{\cm{C}_j}(z_{\cm{C}_j}(k), \mathbf{q}_{\cm{C}_j}(k)) \leq 0, \ \  \forall \cm{C}_j \in \cm{P}(k),  \label{eq:constr_T_2}\\
    &  q_{\cm{C}_j}^\nc{min} \mathbf{1}_{N_\nc{p}} \leq \mathbf{q}_{\cm{C}_j}(k) \leq q_{\cm{C}_j}^\nc{max} \mathbf{1}_{N_\nc{p}},   \ \ \forall \cm{C}_j \in \cm{P}(k),  \\
    & \mathbf{q}_{\cm{C}_0}(k) \geq \mathbf{0}_{N_\nc{p}},\\
    & \sum_{l=0}^{N_\nc{cl}(k)} \mathbf{q}_{\cm{C}_l}(k) =  Q_\nc{T} \mathbf{1}_{N_\nc{p}}, \label{eq:alpha_loops_3}
\end{align}
\end{subequations}
%\forall j \in \{1, ..., N_\nc{cl}(k)\}
with $\mathbf{q}_{\cm{C}_0}(k)$ being the flow surplus over~$Q_\nc{T}$ that the agents decide not to use. Likewise, $f_{\cm{C}_0}(\mathbf{q}_{\cm{C}_0}(k)))$ is a (possibly nonzero) cost associated with sending flow to the sink loop. %, which in turn can be set different from zero.  

\subsection{Distributed coordination using ALADIN}\label{sec:aladin_alg}
Problem~\eqref{eq:loops_problem_3} is an optimal resource allocation problem of the form of those that can be solved in a distributed manner by implementing ALADIN. This algorithm involves an iterative procedure that is briefly introduced below. In this regard, let subscript $p$ enumerate the iterations, $\lambda$ be the multiplier associated with constraint~\eqref{eq:alpha_loops_3}, and consider some time step $k \in \{0, \delta^\nc{c}, 2\delta^\nc{c}, ...\}$. Also, consider  a positive definite scaling matrix~$\Sigma$,  a termination tolerance~$\epsilon$, an initial guess for the primal variables~$\mathbf{y}^{0}=[\mathbf{y}_{\cm{C}_j}^0]_{j=0}^{N_\nc{cl}(k)}$, and some~$\lambda^0$, $\mu^0>0$, and $\rho^0>0$. Then, flow sequences $\mathbf{q}_{\cm{C}_j}(k)$ for all $\cm{C}_j$ are computed by implementing the following steps starting from~$p=0$.  See~\cite{houska2016augmented} and \cite{engelmann2022aladinalpha} for further details.
\begin{itemize}[itemsep=2pt]
    \item[1.] \textit{Parallelizable decentralized step}: All agents  $j \in \{1,2,...,N_{\nc{cl}}\}$ solve locally the following decoupled nonlinear problem: %\footnote{For the sake of clarity, let us omit here time index $k$.} 
        \begin{subequations}\label{eq:NLP_parallel}
\begin{align}
    \min_{\mathbf{q}_{\cm{C}_j}} &  \ f_{\cm{C}_j}(z_{\cm{C}_j}, \mathbf{T}^\nc{ref},  \mathbf{q}_{\cm{C}_j}) + (\lambda^p)^\top \mathbf{q}_{\cm{C}_j} \!+\!\dfrac{\rho^p}{2}\Vert \mathbf{q}_{\cm{C}_j} - \mathbf{y}_{\cm{C}_j}^p\Vert^2_{\Sigma} \nonumber\\
    \text{s.t.} &\ \ 
     h_{\cm{C}_j}(z_{\cm{C}_j}, \mathbf{q}_{\cm{C}_j}) \leq 0, \\ & \ \ q^\nc{min}_{\cm{C}_j} \mathbf{1}_{N_\nc{p}} \leq \mathbf{q}_i \leq q^\nc{max}_{\cm{C}_j}  \mathbf{1}_{N_\nc{p}},
\end{align}
\end{subequations}
\noindent where, for clarity, we have omitted time index~$k$. For the sink artificial loop, we consider additional agent $j=0$, which solves a similar problem considering~$f_{\cm{C}_0}(\mathbf{q}_{\cm{C}_0})$.
    \item[2.]  Let $\mathbf{q}_{\cm{C}_j}^{p}$ be the solution of \eqref{eq:NLP_parallel} for the $j$-th cluster. Then, if $\Vert \sum_{j=0}^{N_\nc{cl}} \mathbf{q}_{\cm{C}_j}^p- Q_\nc{T} \Vert \leq \epsilon$ and $\Vert \sum_{j=0}^{N_\nc{cl}} 
 (\mathbf{q}_{\cm{C}_j}^p - \mathbf{y}_{\cm{C}_j}^p) \Vert \leq \epsilon$, exit the algorithm.
    \item[3.] \textit{Sensitivity evaluations}: All agents $j$ compute gradients $g_i^p = \nabla f_{\cm{C}_j}(\cdot)$, a positive definite Hessian approximation~$H_{\cm{C}_j}^p$, and constraints Jacobian~$G_{\cm{C}_j}^p$~\cite{houska2016augmented}.
    \item[4.] \textit{Coordination step}: Solve the following overall quadratic program~(QP):
    \begin{subequations}\label{eq:QP_ALADIN}
        \begin{align} 
                \min_{s, \Delta \mathbf{q}} &\     \sum_{j=0}^{N_\nc{cl}} \!\left(\dfrac{1}{2} \Vert \Delta\mathbf{q}_{\cm{C}_j}\Vert_{H_{\cm{C}_j}^p}^2\! + (g_{\cm{C}_j}^p)^\top \Delta \mathbf{q}_{\cm{C}_j}\!\right) + r(s, \lambda^p, \mu^p)\nonumber \\
    \text{s.t.} &\ \ 
     \sum_{j=0}^{N_\nc{cl}}  (\mathbf{q}_{\cm{C}_j}^p+ \Delta \mathbf{q}_{\cm{C}_j}) -  Q_\nc{T} = s, \label{eq:constraint_coordination_QP}\\
     & \ \ G_{\cm{C}_j}^p \Delta \mathbf{q}_{\cm{C}_j} = 0, \forall {\cm{C}_j} \in \cm{P}.
     \end{align}
         \end{subequations}
         \noindent where $\Delta \mathbf{q} = [\Delta \mathbf{q}_{\cm{C}_j}]_{j=0}^{N_\nc{cl}}$ and~\mbox{$r(\cdot) = {\lambda^p}^\top s\!+\!\mu^p/2\Vert s\Vert^2$}.
    \item[5.] Finally, update the primal and dual variables as follows:
    \begin{equation}
    \begin{split}
        &\mathbf{y}^{p+1} = \mathbf{y}^{p} + \beta_1^p(\mathbf{q}^{p} - \mathbf{y}^{p}) + \beta_2^p \Delta\mathbf{q}^{p}, \\
        &\lambda^{p+1} = \lambda^{p} + \beta_3^p(\lambda_\nc{QP}^{p} - \lambda^p),
        \end{split}
    \end{equation}
    \noindent where $\mathbf{q}^{p} = [\mathbf{q}_{\cm{C}_j}^{p}]_{j=0}^{N_\nc{cl}}$, $\Delta\mathbf{q}^{p}$ is obtained from the solution of~\eqref{eq:QP_ALADIN}, and~$\lambda_\nc{QP}^{p}$ is the multiplier associated with constraint \eqref{eq:constraint_coordination_QP}. Likewise, factors $\beta_1$, $\beta_2$, and $\beta_3$ are computed following~\cite{houska2016augmented}.
\end{itemize}

\begin{remark}
    %The coordination step of ALADIN involves centralized QP~\eqref{eq:QP_ALADIN}, which in turn considers the solution of~\eqref{eq:NLP_parallel}. 
    The QP in the coordination step could be solved by the supervisor after communicating with the agents, or in a distributed manner using \textit{bi-level} ALADIN~\cite{engelmann2020decomposition}.
\end{remark}

\subsection{Pseudocode}\label{sec:control_alg}

Finally, the pseudocode of the proposed algorithm is summarized in Algorithm~\ref{alg:cap}. Recall that $\Delta t^\nc{c}$ is the control time step and that the system is simulated using a discrete-time version of~\eqref{eq:loops_model} for all $i\in\cm{N}$, where the integration step size is $\Delta t^\nc{s}$. Likewise, the inlet temperature dynamics are modeled considering  the following transfer function: 
\begin{equation}\label{eq:inletT}
    \dfrac{T^\nc{in}(s)}{T^\nc{out}(s)-80}= \dfrac{1}{600s +1},
\end{equation}
where $T^\nc{out}$ is the overall outlet temperature of the solar field, and $T^\nc{out}-80$ºC approximates the outlet temperature of the steam generator. In this regard, for all instants $k$, we consider $T^\nc{out}(k) =\sum_{i\in \cm{N}} q_i(k-1) T_i^\nc{out}(k)/\sum_{i \in \cm{N}} q_i(k-1)$.

\begin{algorithm}
\caption{Control algorithm}\label{alg:cap}
Define a maximum number of clusters $N_\nc{cl}^\nc{max}$, an initial partition $\cm{P}(0) \!=\!\{\cm{C}_1, ..., \cm{C}_{N_\nc{cl}(0)}\}$, with~\mbox{$N_{\nc{cl}}(0)  \leq N_\nc{cl}^\nc{max}$},~and let the partition 
 be updated every~$\Delta t^\nc{cl}\!=\!\delta^\nc{cl} \Delta t^\nc{s}$. Also, assign each $\cm{C}_j$ to agent $j$, and the sink loop to agent $0$.  Then, at all instants $k$, proceed \mbox{as follows}:
\begin{algorithmic}[1]
%\REQUIRE $n \geq 0$
\IF{$k \in  \{0, \delta^\nc{c}, 2\delta^\nc{c}, ...\}$}
\IF{$k \in \{\delta^\nc{cl}, 2\delta^\nc{cl}, ...\}$}
\STATE Update the clusters as described in Section~\ref{sec:cluster_formation} and define a partition $\cm{P}(k)= \{\cm{C}_1, ..., \cm{C}_{N_\nc{cl}(k)}\}$ such that $N^\nc{cl}(k) \leq N_\nc{cl}^\nc{max}$.
\ELSE
\STATE Set $\cm{P}(k) \leftarrow \cm{P}(k-1)$.
   % \STATE The supervisor solves problem \eqref{eq:clusters_optimization}, and updates scale factors~$\alpha_{\cm{C}_j}(\cdot)$ for all $\cm{C}_j \in \cm{P}(k)$ according to~\eqref{eq:alpha}.
\ENDIF
\STATE All MPC agents $j \in \{0,2,..., N^\nc{cl}(k) \}$ solve problem~\eqref{eq:clusters_optimization} in a distributed manner by using ALADIN algorithm as described in Section~\ref{sec:aladin_alg}. As a solution, the agents find the flow rates to be pumped to each cluster $\cm{C}_j$ during interval $[k, k+\delta^\nc{c})$.
\STATE For each cluster $\cm{C}_j$, define $q_i(t) = q_{\cm{C}_j}^*(k|k)/|\cm{C}_j|$ for all $i \in \cm{C}_j$ and $t \in [k, k+\delta^\nc{c})$.
\ENDIF
\STATE Simulate the loops dynamics considering~\eqref{eq:loops_model},~\eqref{eq:parameters_loops},~\eqref{eq:inletT}, and the current flow rates for all loop~\mbox{$i \in \cm{N}$}. 
\STATE Set $k \leftarrow k+1$.
\end{algorithmic}
\end{algorithm}

\section{Simulation results}\label{sec:simulations}

In this section, we simulate Algorithm~\ref{alg:cap} on a 10-loop  solar parabolic plant using different values of  $N_\nc{cl}^\nc{max}$ and~$\Delta t^\nc{cl}$, and considering the parameters in Table~\ref{tab:param_sim}. All simulations were carried out in a~1.8 GHz Intel$^\text{\textregistered}$ Core$^\nc{TM}$~i7/16GB RAM computer using Matlab$^\text{\textregistered}$, software CasADi~\cite{Andersson2018}, and toolbox \mbox{ALADIN-$\alpha$}~\cite{engelmann2022aladinalpha}. Also, we used \texttt{ipopt} and \texttt{MA57} for solving \eqref{eq:NLP_parallel} and~\eqref{eq:QP_ALADIN}, respectively. The partitions were found using  function \texttt{kmeans} with the Calinski-Harabasz index~\cite{calinski1974dendrite} defining the~score. 

As a  reference, the results are compared with those obtained considering statically the \textit{finest} and \textit{coarsest} partition of the system. The former corresponds to running Algorithm~\ref{alg:cap} with initial singleton partition $\cm{P}(0) = \{\{1\}, \{2\}, \hdots, \{10\}\}$  and $\Delta t^\nc{cl} = \infty$. By contrast, the latter corresponds to $\cm{P}(0) = \{1,2,..., 10\}$ and $\Delta t^\nc{cl} = \infty$, i.e., a single controller uses a lumped parameter model of the entire solar field and distributes equally the flow among all loops. For the sake of clarity, these two approaches will be denoted as $\text{DMPC}_\nc{fin}$ and $\text{MPC}_\nc{coar}$, respectively.

% To  benchmark the obtained results, we consider the following approaches: 
% \begin{itemize}
%     \item[(i)] \textit{Cooperative DMPC based on the singleton  partition}:  In this case, each agent controls a single loop using prediction model~\eqref{}. 
%     %This approach leads to a distributed resolution of problem~\eqref{eq:cen_problem} at all control time steps. 
%     %Each agent $i \in \{\{1\}, \{2\}, \hdots, \{10\}\}$ considers a local model of the $i$-th loop given by~\eqref{eq:model_opt_problem_bottom}. 
% \item[(ii)] \textit{Centralized MPC based on static partition $\cm{P} = \cm{N} = \{1,2,..., 10\}$}: That is, a single controller 

% with a lumped parameter model of the entire solar field: In this case, a single controller sets dynamically an uniform flow rate distribution for all loops. This case corresponds to solving problem~\eqref{eq:clusters_optimization} using static partition $\cm{P} = \cm{N} = \{1,2,..., 10\}$. 
% % \item[(iii)] \textit{Decentralized MPC}: That is,~$\cm{P}(k)=\{\{1\}, \{2\}, \hdots, \{10\}\}$ and~\mbox{$\alpha_{\{i\}}=1/10$} for all~$k\geq 0$ and $i \in \{1,2,..., 10\}$. 
% % \item[(iv)] \textit{No-valves centralized MPC:} The HTF flow is distributed equally among all loops by a single controller that considers all loops dynamics. 
% \end{itemize} 

\begin{table}[b]
    \centering
    \caption{Parameters used in the simulations}
    \begin{tabular}{lll|lll} %p{0.75cm} p{1.15cm} p{0.5cm}|p{0.6cm} p{2.2cm} p{0.35cm}}
     & Value & Unit &  & Value & Unit \\ \hline
    $q^\nc{min}$ & 0.2$\cdot 10^{-3}$ & m$^3$/s & $\Delta t^\nc{s}$ & 0.5 & s  \\ 
    $q^\nc{max}$ & 2$\cdot 10^{-3}$ & m$^3$/s  &   $\Delta t^\nc{c}$ & 30 & s  \\
    $T^\nc{min}$ & 220 & \degree C & $w_\nc{e}$ & 1$\cdot 10^{-3}$ & \\ 
    $T^\nc{max}$ & 305 & \degree C & $w_\nc{q}$ & 1 & -    \\
    $A$ & 5.067$\cdot 10^{-4}$ & m$^2$  & $N_\nc{p}$ & 5 & -      \\
    $L$ & 142 & m &  $\epsilon$ & $1 \cdot 10^{-5}$ & -  \\ 
    $S$ & 267.4 & m$^2$ &   $Q_\nc{T}$ &   $9$ & l/s
    \end{tabular}
    \label{tab:param_sim}
\end{table}

\begin{figure}[t]
	\centering
		\includegraphics[scale=.55,trim={2.5cm 7.6cm 3cm 8cm},clip]{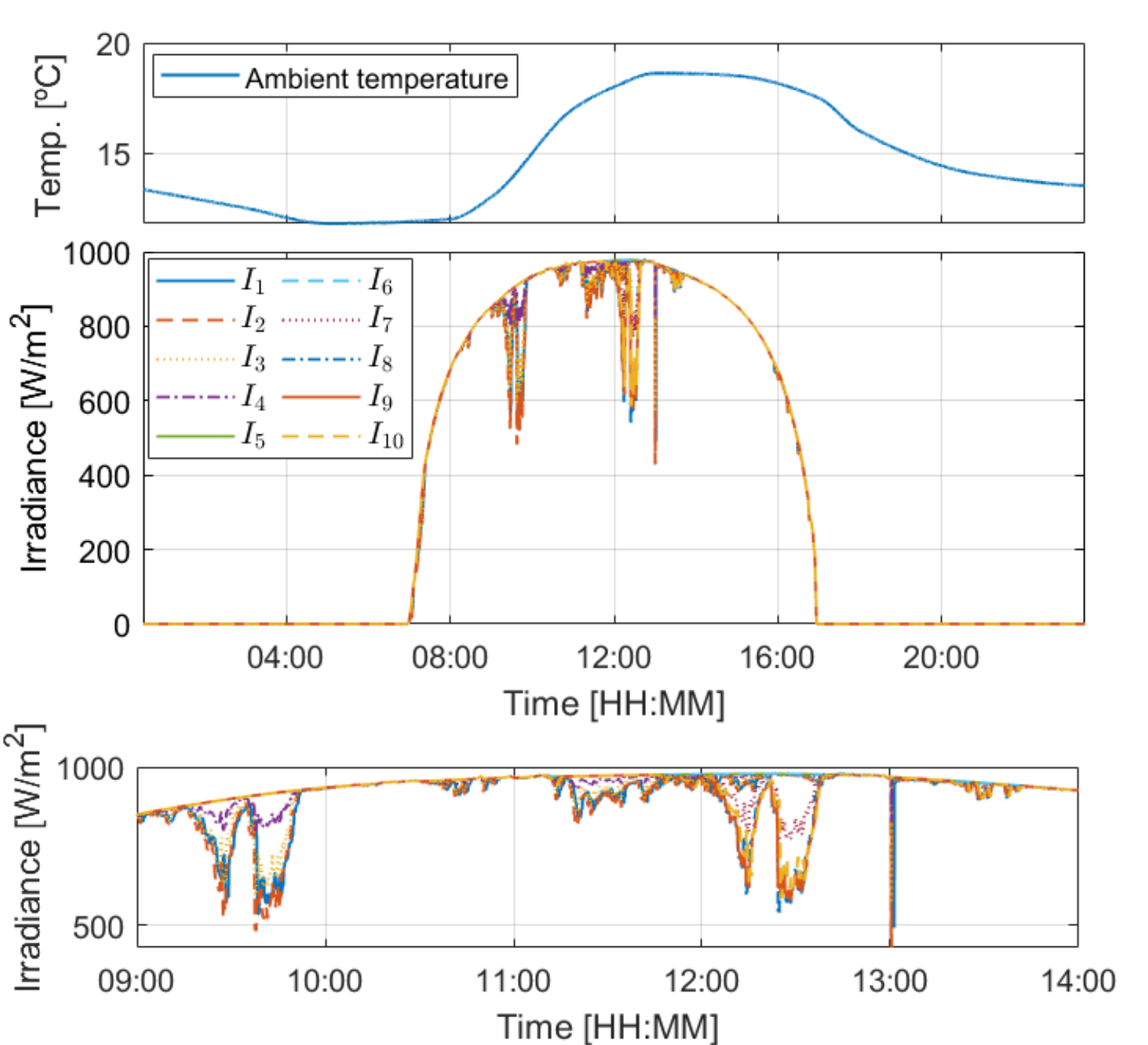}
  \vspace{-2pt}
	\caption{Evolution of the ambient temperature and of the direct normal irradiance for every loop. The bottom plot zooms the irradiance graph in a period affected by clouds.}
	\label{fig:irradiance}
\end{figure}

\begin{figure}[!t]
 \centering
     \begin{subfigure}[t]{0.48\textwidth}
         \centering
         \includegraphics[scale=.55,trim={3cm 9.3cm 1.5cm 9.3cm},clip]{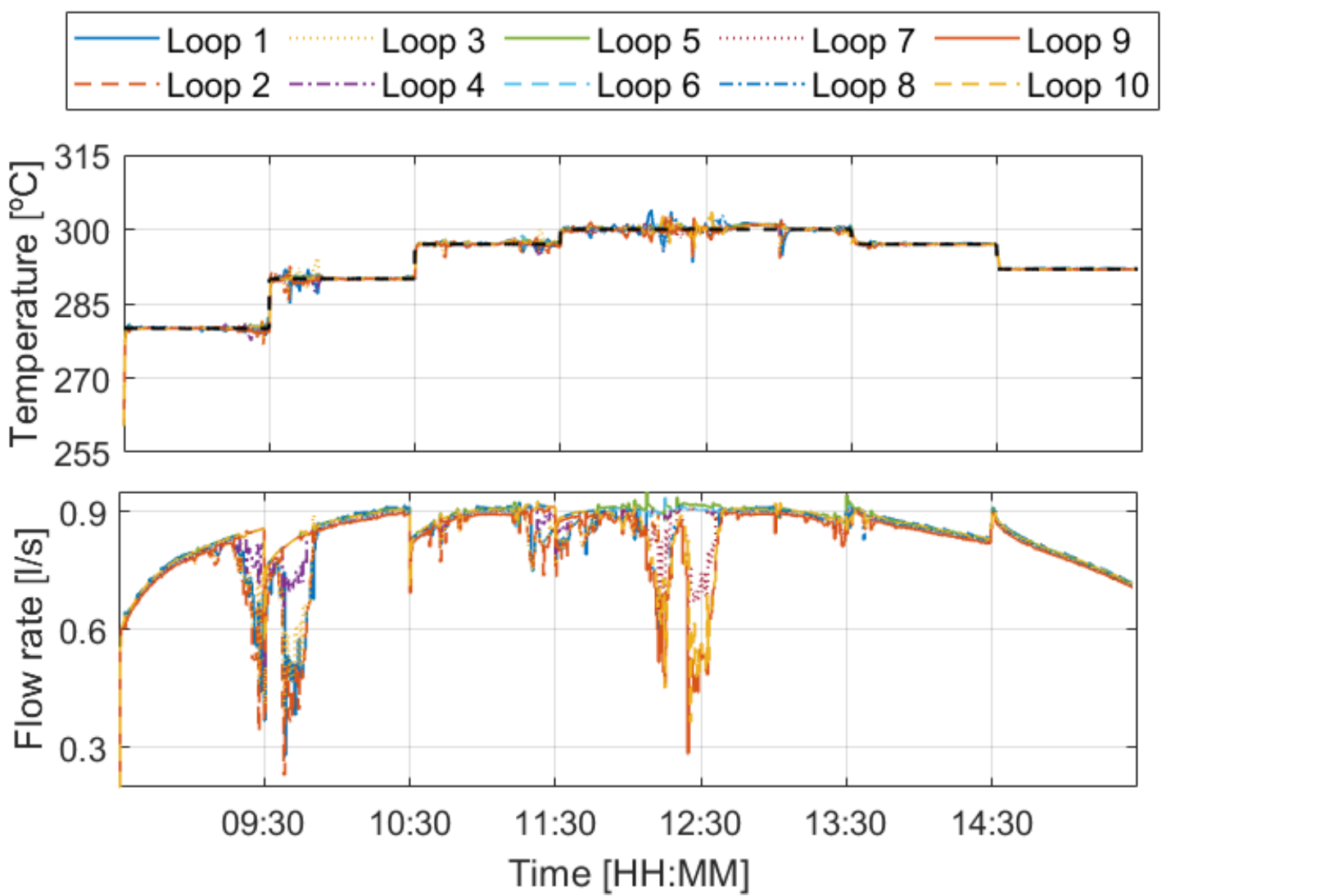}\label{fig:clust_Tq}
         \vspace{-0.4cm}
         \caption{Proposed DMPC with  $N_\nc{cl}^\nc{max}=5$ and $\Delta t^\nc{cl} = 2.5$ min }
     \end{subfigure} 
      \begin{subfigure}[t]{0.48\textwidth}
         \centering
         \includegraphics[scale=.55,trim={3cm 9.3cm 1.5cm 8.5cm},clip]{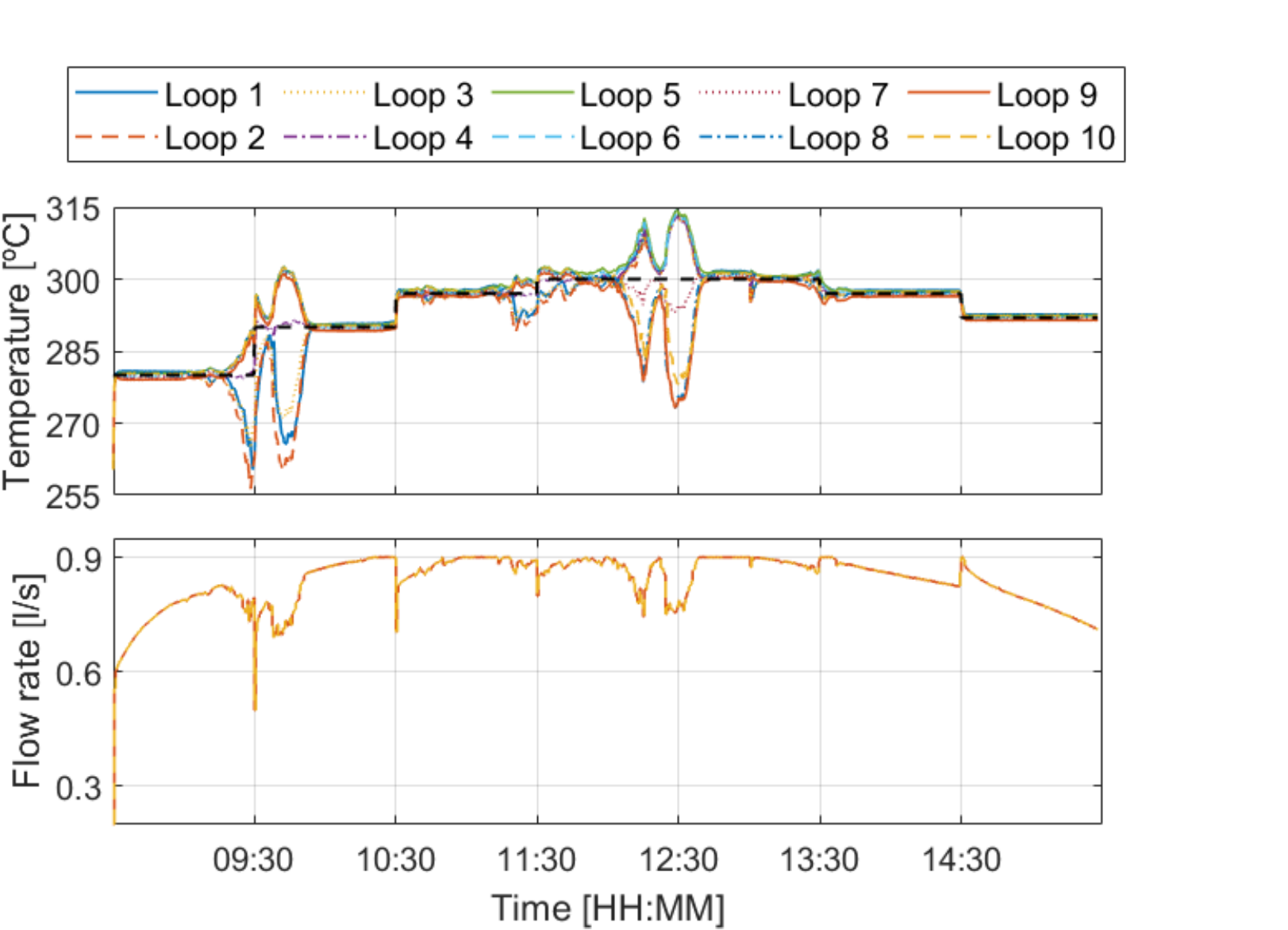}\label{fig:no_valves_Tq}
         \vspace{-0.5cm}
         \caption{$\text{MPC}_\nc{coar}$}
     \end{subfigure}
	\caption{Evolution of the loops outlet temperature and of the HTF flow rates with different controllers. The dashed black line indicates the reference temperature.}% In the temperature plots, the blue line shows the outlet temperature of the field, whereas the grey dotted lines indicate the maximum and minimum outlet temperature among all~loops. }
	\label{fig:Tq}
\end{figure}

The simulations consider a 7 hours period (8:30am-3:30pm) of a cloudy day in which the irradiance and ambient temperature evolve as shown in Fig.~\ref{fig:irradiance}.  The outlet temperatures and flow rates evolution is illustrated in Fig.~\ref{fig:Tq}~(a) for the case of $N_\nc{cl}^\nc{max}\!=\!5$ and~$\Delta t^\nc{cl}\!=\!2.5$ min. As can be seen, the loops outlet temperatures follow closely the reference, and the flows decrease as the irrandiance falls.  However, the system performance underwent a significant deterioration when using $\text{MPC}_\nc{coar}$ (see Fig.~\ref{fig:Tq}~(b)). Note that in the latter case all loops receive the same flow, and hence there is no chance of adjusting it to the space-varying conditions in the solar field. Particularly, given~\eqref{eq:alloc}, the maximum flow that the loops can get with $\text{MPC}_\nc{coar}$ is $Q_\nc{T}/10=0.9$ l/s, whereas in the DMPC case we obtained $\max_{i,k} q_i(k)=0.95$ l/s.  Also, when the overall outlet temperature approaches~$T^\nc{max}$, controller $\text{MPC}_\nc{coar}$ increases the flow in all loops, and this decreases the temperature even of those that were already below the reference.

\renewcommand{\arraystretch}{1.35}

\renewcommand{\arraystretch}{1.2}
\begin{table}[b] 
\caption{Cumulative performance costs and clusters size}
    \centering 
    \begin{tabular}{cccccccc}
     & & & $J_\nc{cum}$ & $\bar{e}$ & \makecell{Mean no. of\\ loops/cluster} \\ \hline
  \multirow{2}{*}{\rotatebox[origin=c]{90}{\makecell{Static \\part.}}} & \multicolumn{2}{c}{DMPC$_\text{fin}$} & 172.08 & 9.84 & 1 \\
  & \multicolumn{2}{c}{MPC$_\text{coar}$} &  8.22$\cdot 10^3$ & 29.83 &  10 \\ \hline \hline %\vspace{2pt}
  & $N_\nc{cl}^\nc{max}$ & $\Delta t^\nc{cl}$ [min] & & & \\
      \multirow{4}{*}{\rotatebox[origin=c]{90}{\makecell{Time-varying \ \ \\partition}}}  
      & 8 & 1.5& 176.89 & 9.86  & 1.39\\ 
      & 6 & 1.5 &     195.02  & 10.23 & 2.03 \\
      & 5 & 1.5 &  215.44 & 10.60 & 2.51 \\
      & 5 & 2.5 & 228.59 & 10.63 & 2.53 \\
      & 3 & 5.0 & 749.88 & 19.28 & 4.15 
    \end{tabular}
    \vspace{0.1cm}
    \label{tab:cum_cost}
\end{table}

The system performance is also numerically compared in Table~\ref{tab:cum_cost}, which provides the cumulative costs in different simulations,~i.e., $ J_\nc{cum} = \sum_{k \in \cm{K}} \sum_{i=1}^{10} \left(w_\nc{e}e_i^2(k) + w_\nc{q} q_i^2(k) \right)$, 
% \begin{equation*}
%     J_\nc{cum} = \sum_{k \in \cm{K}} \sum_{i=1}^{10} \left(w_\nc{e}e_i^2(k) + w_\nc{q} q_i^2(k) \right)\!, 
% \end{equation*}
together with the maximum incurred temperature errors, i.e., 
\begin{equation*}
     \bar{e} = \!\max_{\substack{k \in \tilde{\cm{K}}, \\ \ {i\in\{1,..., 10\}}}} |e_i(k)| = \! \max_{\substack{k \in \tilde{\cm{K}}, \\ \ {i\in\{1,..., 10\}}}} 
 |T^\nc{out}_i(k)- T^\nc{ref}(k)|.
\end{equation*}
\noindent Above, $\cm{K}$ represents the set of all simulated time instants, and~$\tilde{\cm{K}} \subset \cm{K}$ contains the instants after the first simulated~5 minutes. Note that set $\tilde{\cm{K}}$ is used not to account for the errors at the beginning of the simulations, which are mainly influenced by the choice of the initial state. In addition, Table~\ref{tab:cum_cost} indicates the mean number of loops per cluster. As expected, finer partitions and reduced~$\Delta t^\nc{cl}$ resulted both in lower performance costs and lower temperature errors. In particular, the proposed approach with $N_\nc{cl}^\nc{max} = 8$ and~$\Delta t^\nc{cl} = 1.5$ min performed comparably to~$\text{DMPC}_\text{fin}$. Likewise, significant improvements with regard to~$\text{MPC}_\text{coar}$ were observed even with only three clusters. Note also that the temperature errors could be reduced if accurate irradiance estimations are available.

\begin{figure}[t]
	\centering
		\includegraphics[scale=.42,trim={0cm 9.8cm 0.5cm 8.8cm},clip]{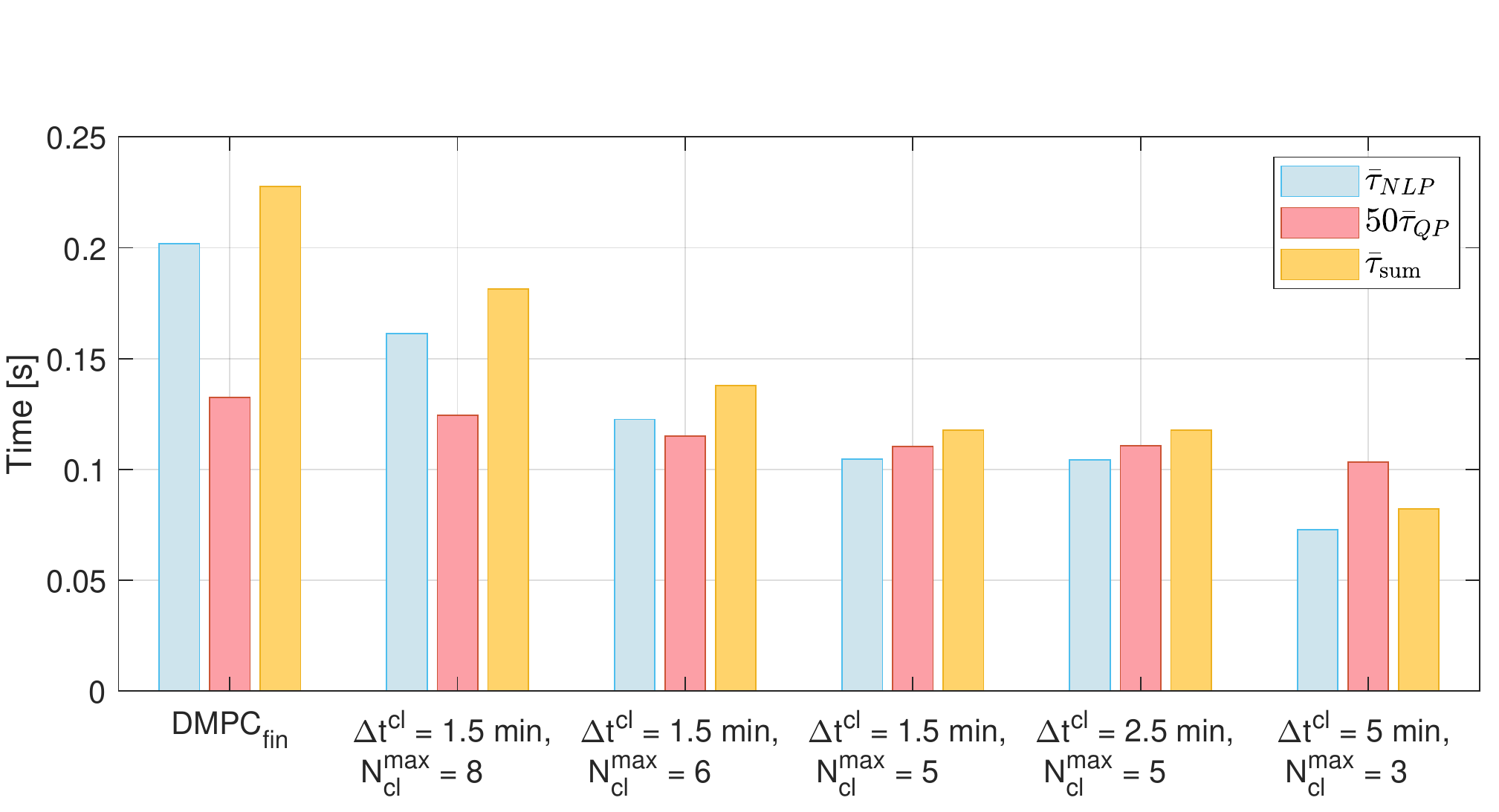}
  %\vspace{-12pt}
	\caption{Computation times of different steps of ALADIN algorithm. For sake of clarity, $\bar{\tau}^\nc{QP}$ is scaled by~50.}
	\label{fig:comp_time}
\end{figure}

Regarding the computation times, Fig.~\ref{fig:comp_time} shows the values of the following indexes:
\begin{equation*}\label{eq:comp_times}
\begin{split}
    &\bar{\tau}^\nc{NLP}\!=\!\dfrac{1}{|\cm{K}^\nc{c}|}\!\sum_{k \in \cm{K}^\nc{c}} \!\!\sum_{j=1}^{N_\nc{cl}(k)} \!\tau^{\nc{NLP}}_{\cm{C}_j} (k),  \ \ \ \bar{\tau}^\nc{QP}\!=\!\dfrac{1}{|\cm{K}^\nc{c}|}\! \sum_{k \in \cm{K}^\nc{c}} \!\tau^{\nc{QP}}(k), \\
    & \bar{\tau}^\nc{sum}\!=\!\bar{\tau}^\nc{NLP}  + \bar{\tau}^\nc{QP} +
 \dfrac{1}{|\cm{K}^\nc{c}|} \sum_{k \in \cm{K}^\nc{c}} \sum_{j=1}^{N_\nc{cl}(k)}  \tau^{\nc{sens}}_{\cm{C}_j}(k), 
    \end{split}
\end{equation*}
which are associated with  different steps of the ALADIN algorithm. Above, $\tau^{\nc{NLP}}_{\cm{C}_j}(k)$ and $\tau^{\nc{sens}}_{\cm{C}_j}(k)$ denote respectively the time spent by agent $j$ solving nonlinear problem~\eqref{eq:NLP_parallel} and computing the sensitives at time step $k$. In addition, $\tau^{\nc{QP}}(k)$ refers to the time spent solving QP problem~\eqref{eq:QP_ALADIN}\footnote{These values were obtained using \texttt{timers.NLPtotTime},  \texttt{timers.QPtotTime} and  \texttt{timers.sensEvalT}, where struct \texttt{timers} is given by ALADIN-$\alpha$ toolbox.}, and~$\cm{K}^\nc{c}$ is the set of instants in which the flow rates are updated. As reflected in Fig.~\ref{fig:comp_time}, finer partitions involve a greater number of variables to coordinate, and led to higher computation times. Notice also that, although steps~1 and 3 of ALADIN can be performed in parallel, increasing the number of distributed agents also demands greater communication links. %Finally, we also observed notable differences in the number of iterations executed by ALADIN. %For example, in the simulation with $N_\nc{cl}^\nc{max}=8$, the mean number of iterations was 8.52, whereas for $N_\nc{cl}^\nc{max}=6$ it was 6.29, and for $N_\nc{cl}^\nc{max}=3$ it resulted in 3.54. 

% Finally, Table~\ref{tab:clust} focuses on the loops clusters formation. In particular, it provides the mean number of clusters and mean number of loops loops per cluster for different setting of $N_\nc{cl}^\nc{max}$ and $\Delta t^\nc{cl}$. As can be seen, a greater number of clusters usually involves a reduced number of loops per cluster, which in turn allows for more accurate predictions and reduced performance costs. 

\section{Conclusions}\label{sec:conclusions}

 A DMPC with time-varying partitioning for optimizing the~HTF flow rates in solar parabolic trough plants has been presented. In this regard, clustering methods are considered for dynamically partitioning the solar field into clusters of similar loops, which are subsequently assigned to a set of MPC agents. The article formulates the associated DMPC problem so that it can be addressed implementing ALADIN algorithm, and illustrates its effectiveness via simulations. In particular, it is shown that the proposed approach can closely approximate that of a DMPC with static finer partitions while reducing the number of variables to be coordinated. 
Future research will include a comparison with ADMM, as well as exploring bi-level ALADIN. Also, we will extend our results to larger plants, and consider the optimization of the setpoint so as to maximize the net electricity production.

% In view of Fig.~\ref{fig:Tq} and Table~\ref{tab:cum_cost}, it should also be noted that the centralized and coalitional controllers introduce further flexibility in the distribution of the HTF among the loops, which allows for a better performance considering the same~$Q_\nc{T}$. 

\bibliography{biblio.bib}
\bibliographystyle{IEEEtran}
\end{document}